\documentclass{sig-alternate}
\usepackage{verbatim} 
\usepackage{graphicx}
\usepackage{url}
\begin{document}
\conferenceinfo{USEWOD2011}{Hyderabad, India}

\title{An Empirical Study of Real-World SPARQL Queries}

\numberofauthors{4} 
%
\author{
%
%
\alignauthor
Mario Arias Gallego\\
       \affaddr{Computer Science Dept.}\\
       \affaddr{Univ. of Valladolid, Spain}\\
       \email{mario.arias@gmail.com}
\alignauthor
Javier D. Fern\'andez\\
       \affaddr{Computer Science Dept.}\\
       \affaddr{Univ. of Valladolid, Spain}\\
       \email{jfergar@infor.uva.es}
\alignauthor
Miguel A. Mart\'inez-Prieto\\
       \affaddr{Computer Science Dept.}\\
       \affaddr{Univ. of Valladolid, Spain \&}\\
       \affaddr{Univ. of Chile, Chile}\\
       \email{migumar2@infor.uva.es}
\and  
\alignauthor
Pablo de la Fuente\\
       \affaddr{Compt Science Dept}\\
       \affaddr{Univ. of Valladolid, Spain}\\
       \email{pfuente@infor.uva.es}
}
\date{28 March 2011}

\maketitle
\begin{abstract}
Understanding how users tailor their SPARQL queries is crucial when designing query evaluation engines or fine-tuning RDF stores with performance in mind. In this paper we analyze 3 million real-world SPARQL queries extracted from logs of the DBPedia and SWDF public endpoints. We aim at finding which are the most used language elements both from syntactical and structural perspectives, paying special attention to triple patterns and joins, since they are indeed some of the most expensive SPARQL operations at evaluation phase. We have determined that most of the queries are simple and include few triple patterns and joins, being Subject-Subject, Subject-Object and Object-Object the most common join types. The graph patterns are usually star-shaped and despite triple pattern chains exist, they are generally short.  
\end{abstract}




\section{Introduction}
RDF\footnote{\texttt{\small http://www.w3.org/TR/REC-rdf-syntax/}} provides a simple declarative data model of triples $(subject,predicate,object)$ to describe resources. 
The number of RDF data sets has increased in diverse areas of application such as bioinformatics, social networks, geographic locations, books or films. The \textit{Linked Data Project}\footnote{\tt\small http://linkeddata.org} has emerged as an initiative to promote the use of RDF to publish structured data on the Web in a distributed and interconnected manner~\cite{BHIB:08}. 
Linked Open Data (LOD) cloud estimations\footnote{\tt\small http://www4.wiwiss.fu-berlin.de/lodcloud/} show that more than 25 billion RDF triples are available and interconnected by roughly 1 billion links. 

SPARQL\footnote{\texttt{\small http://www.w3.org/TR/rdf-sparql-query/}} is a declarative language recommended by the W3C for extracting information from RDF graphs. It proposes graph pattern matching facilities to perform searches and data extraction. For instance, it provides the possibility of extracting subgraphs using the \texttt{CONSTRUCT} keyword, or finding certain variable bindings using the \texttt{SELECT} clause. The semantics and complexity of the SPARQL query language have been fairly studied theoretically, showing that SPARQL algebra has the same expressive power as relational algebra~\cite{Perez2009}, although their conversion is not trivial~\cite{Angles2008, Cyganiak2005}. 

Several works~\cite{Neumann2010,Atre2010} explore efficient SPARQL evaluation methods based on query evaluation optimization~\cite{Groppe2009}. Some heuristics include triple pattern reordering based on selectivity estimation~\cite{Stocker2008}, dynamically restricting triple patterns~\cite{Groppe2009}, RISC-style query processing~\cite{Neumann2010} and optimization based on ``star shaped groups''~\cite{Vidal2010}, \emph{i.e.}, different triple patterns around one or few common variables. 
Some techniques focus on minimizing the processing time of joins. In~\cite{Abadi2007}, subject-subject joins are assumed to be very frequent operations, and they can be carried out in linear time (\emph{w.r.t.} the size of the tables). Multi-way joins can also be performed instead of multiple individual joins~\cite{Myung2010}. RDF store benchmarking has also conjectured about SPARQL special features to provide sets of representative queries~\cite{Bizer2009,Schmidt2009}. 

A recent work~\cite{Miller2010} motivates the need of characterizing LOD usage patterns and analyzing who utilizes the information and how. This knowledge helps understand which resources are more useful and allows to adapt LOD repositories to suit the real needs of the users. This study highlights the peculiarities of analyzing LOD web server logs compared to the well-known traditional web log analysis. 
For instance, it compares the proportion of accesses to the two views of the same resource: the traditional human HTML view, and the semantic RDF perspective. 
Ultimately, it performs a basic analysis of a set of SPARQL queries, counting the type of queries and triple patterns.

The main objective of this work is analyzing real-world SPARQL queries to depict what kind of accesses the users perform, focusing on those clauses that are more expensive in terms of query evaluation, both from planification and index access points of view. This study is data-set-independent, in the sense that anyone could follow our methodology to analyze any SPARQL log, querying any RDF data set, supported by any RDF store or query engine. We consider that our study may assist the designers of indices, stores, optimizers and benchmarks in making reasonable assumptions and taking plausible decisions. 

In Section~\ref{s:sla}, we first describe the properties of the logs and the preprocessing steps. Then, we provide a comprehensive analysis of the clauses and structure of the queries. Finally, Section~\ref{s:cfw} summarizes our conclusions and future work.

\section{SPARQL Log Analysis}
\label{s:sla}

\begin{table}
\centering
\scriptsize
\begin{tabular}{|c|r|r|} \hline
 & DBPedia & SWDF \\\hline
Total Queries & 5 166 272 & 2 062 508 \\\hline
Duplicates from same host & 51.7\% & 69.8\% \\\hline
Parse error & 4.37\% & 1.03\% \\\hline
Analyzed & 43.9\% & 29.1\% \\\hline
\end{tabular}
\caption{DBPedia and SWDF query log statistics.}
\label{tab:global}
\end{table}

We use the logs from the USEWOD2011 Challenge~\cite{USEWOD2011}, kindly provided by the organisation. They consist of several months of usage data (server logs) from DBPedia\footnote{\texttt{\small http://www.dbpedia.com}} (about general knowledge) and Semantic Web Dog Food\footnote{\texttt{\small http://data.semanticweb.org}} (about authors and publications). Since we have two sources of information, we analyze them aside and then compare their results. Both query sets are statistically relevant due to the amount and heterogeneity of the users generating them, including both human users and machine agents~\cite{Miller2010}.

We first extract the queries from the HTTP log to their textual representation, obtaining roughly 5 million queries for DBPedia and 2 million for SWDF. Then we parse each query using Jena\footnote{\texttt{\small http://www.openjena.org}} and extract all relevant features using a tool we specifically designed for this task. Users tend to repeat some queries, so lots of them are mere duplicates. Since this fact might bias our results, we exclude from our study all the identical queries generated from the same host, and all those that do not comply with the SPARQL grammar specification and result in parsing errors. Table~\ref{tab:global} summarizes some statistics for the original and resultant data sets.

We first investigate which are the most common types of SPARQL queries. The most frequent one is \texttt{SELECT}, comprising the $96.9\%$ and $99.7\%$ of DBPedia and SWDF queries respectively. It is more surprising that \texttt{ASK} ($1.6\%$/$0.2\%$), \texttt{CONSTRUCT} ($1.5\%$/$0.01\%$) and \texttt{DESCRIBE} ($0.002\%/0.002\%$) are scarcely used. 

\begin{figure}
\centering
\includegraphics[width=7cm]{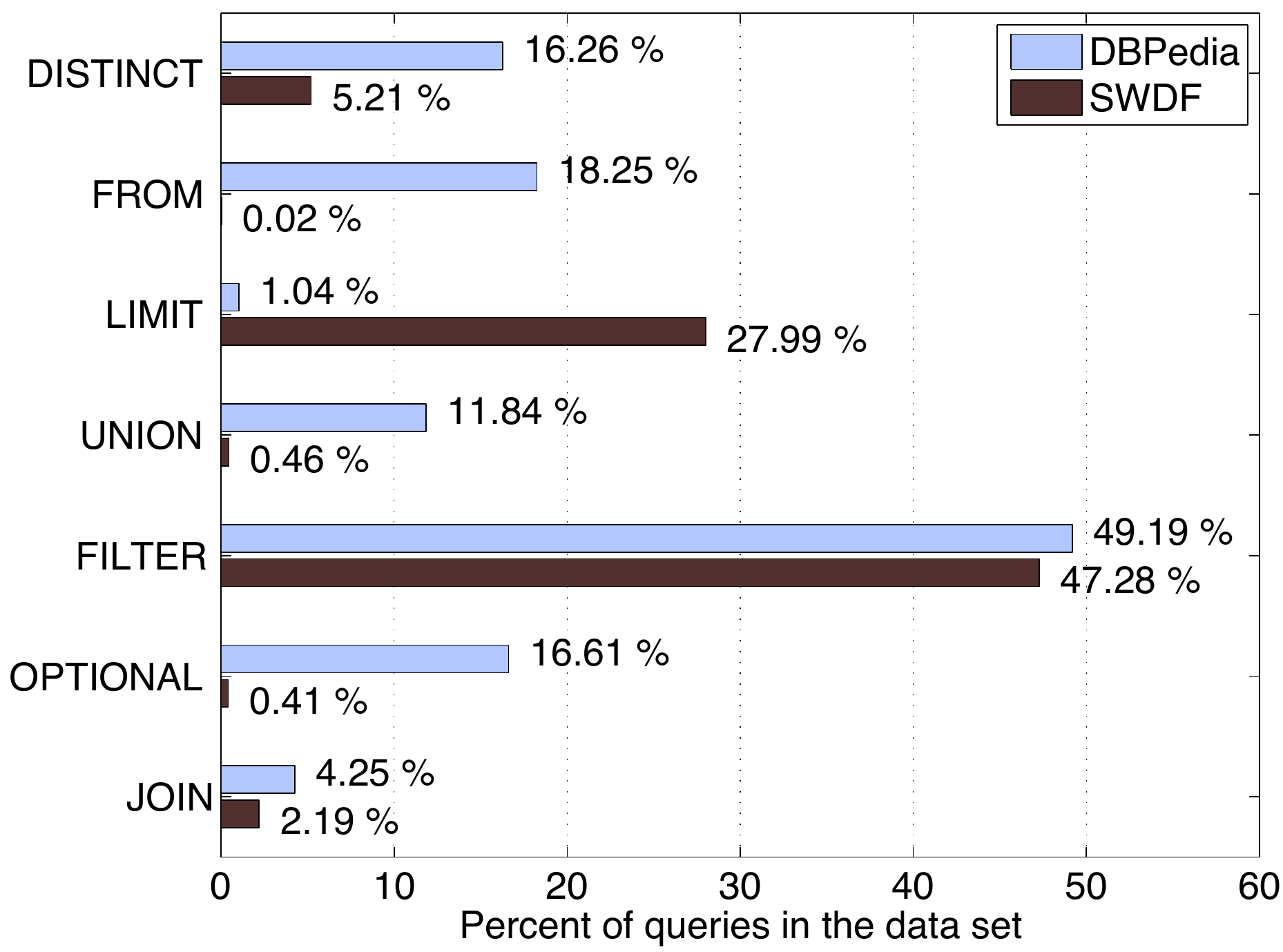}
\caption{Percentage of queries using the different SPARQL features at least once.}
\scriptsize
\label{fig:constructs}
\end{figure}

Then, we estimate the frequency of appearance of the different SPARQL features (figure~\ref{fig:constructs}). \texttt{FILTER} is the most used one by almost $49\%$ in both sources. This is is a relevant result considering that some query planification algorithms~\cite{Vidal2010} \emph{push} filters to be executed first, thus, the space of RDF triples to be explored is considerably reduced. 
\texttt{LANG} is the most used function, it occurs in the $28\%$ of the total DBPedia filters. However, it was not mentioned in SWDF, which is quite obvious considering that DBPedia publishes contents in many languages and SWDF only contains English literals. 
The following most-used function in DBPedia is the \texttt{equal} comparator (23\%) which holds the first position in SWDF (93\%). A further analysis of the filter expressions reveals that 99.4\% of the filters only affect one variable. These facts envisage that adequate indices would enhance access operations on the many queries including filters.

We also analyze the usage of \texttt{DISTINCT}, which ensures that no duplicates are returned. We observe that it is more popular on DBPedia than SWDF, perhaps because of the complexity of its schema. We also study the appearance of \texttt{SELECT REDUCED} that lets the SPARQL engine remove duplicates if possible, but not mandatory. We discover that only an insignificant amount of two queries did use this method, therefore RDF store designers should not rely on users using this modifier. The \texttt{FROM} feature is widely used on DBPedia, which is composed by several data sources, but it is not used on SWDF since it only has one big graph and this would be redundant. We also note that there is a lack of usage of the features \texttt{ORDER BY}, \texttt{GRAPH}, \texttt{FROM NAMED} and \texttt{OFFSET} which occurred less than $0.5\%$ in our tests. 

Some authors~\cite{Neumann2010} state that the most used feature in SPARQL is conjunction. While this statement holds true, the amount of disjunctions (\texttt{UNION}) is not small enough to be overlooked whatsoever, since it appears in $11.84\%$ of DBPedia queries. We also find that there is a significative amount of \texttt{OPTIONAL} blocks. This result is critical, because some studies proved that the optional operator from the SPARQL algebra is the major culprit of the query evaluation being PSPACE-complete~\cite{Perez2009}. 

\begin{table}
\centering
\scriptsize
\begin{tabular}{|c|r|r|} \hline
{\bf Pattern} & {\bf DBPedia} & {\bf SWDF} \\\hline
C C V & 66.35\% & 47.79\%\\\hline
C V V & 21.56\% & 0.52\%\\\hline
V C C & 7.00\% & 46.08\%\\\hline
V C V & 3.45\% & 4.21\%\\\hline
C C C & 1.01\% & 0.001\%\\\hline
V V C & 0.37\% & 0.19\%\\\hline
C V C & 0.20\% & 0.006\%\\\hline
V V V & 0.04\% & 1.18\%\\\hline
\end{tabular}
\caption{Triple Patterns {\small (C=Constant, V=Variable)}.}
\label{tab:triplepattern}
\end{table}

Once we have shown a basic insight on the usage of single elements, we proceed to perform a higher level analysis on the structure of the query expressions. SPARQL provides a means to match graph patterns by specifying several \emph{triple patterns}, \emph{i.e.} RDF triples in which each element can be a variable. Our first objective is checking which ones are the most frequent ones (table~\ref{tab:triplepattern}). We noticed that \texttt{C C V} (\emph{i.e.} given a subject and a predicate, obtain the value) is the most used one. \texttt{C V V} is also very common, which means given a subject, obtain all different properties and their values. The third most used pattern is \texttt{V C C} which obtains all subjects with a given property and value. A comparison of DBPedia and SWDF shows a significant difference suggesting that the usage of triple patterns is highly-dependent on the kind of information provided and its structure. These results are also very valuable when choosing indexing schemas. Based on the results shown in table~\ref{tab:triplepattern}, since \texttt{C C V} is a very common access operation, we can foresee that a multifield index on (Subject-Predicate) would significantly improve search performance. 

\begin{figure}
\centering
\includegraphics[width=7cm]{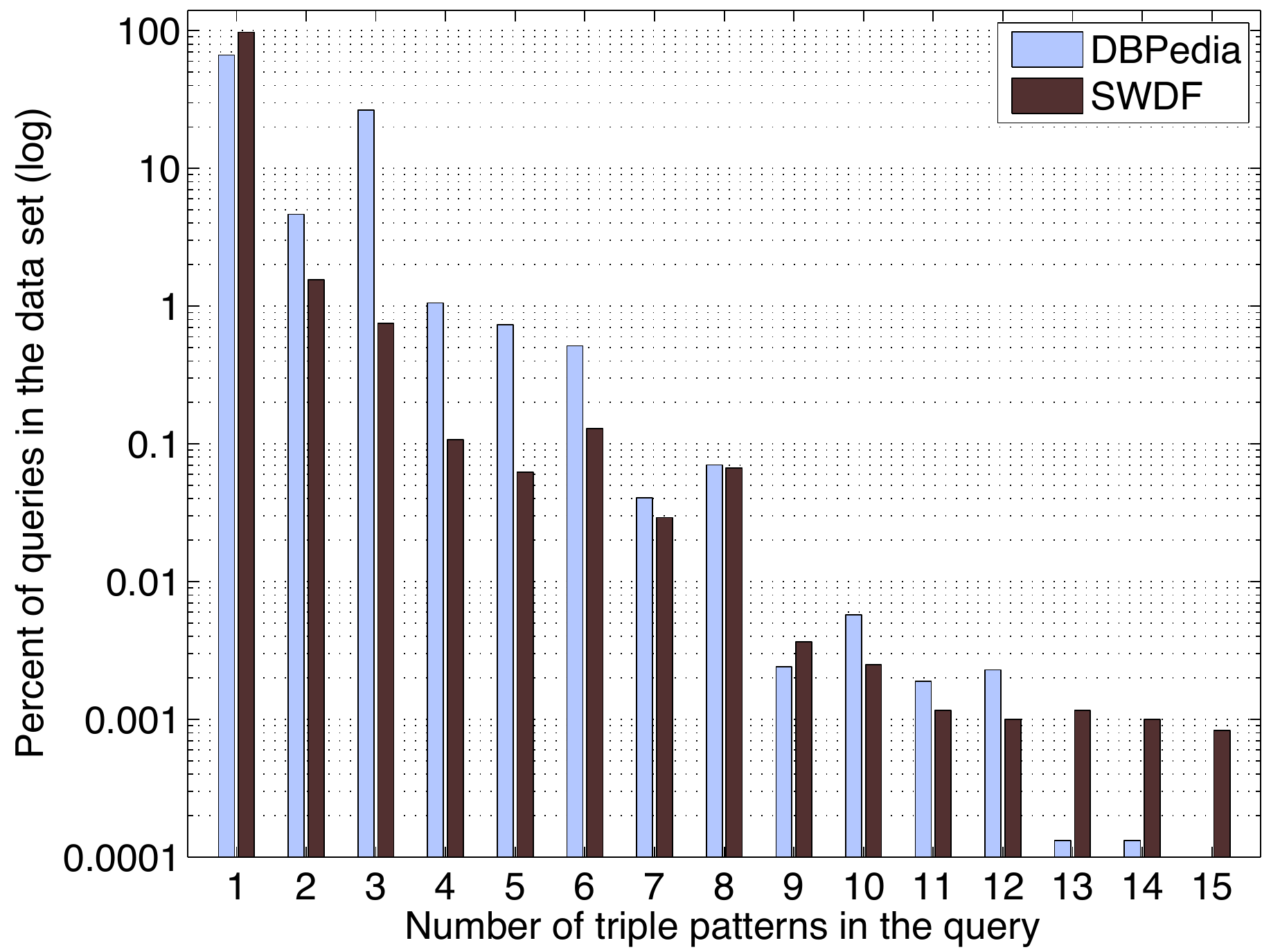}
\caption{Percentage of triple patterns per Query.}
\scriptsize
\label{fig:triplepattquery}
\end{figure}

Having completed the study of the simple patterns, we need to investigate how they blend together. The first obvious question is how many triple patterns appear in each query (figure~\ref{fig:triplepattquery}). We see that most of the queries contain one single triple pattern ($66.41\%$ in DBPedia, $97.25\%$ in SWDF). Thereafter they follow the rule that most queries have few triple patterns and fewer queries have many triple patterns. Note that the figure is in logarithmic scale, so the number of queries with two patterns is one order of magnitude less than those with one pattern for DBPedia, and almost two orders of magnitude for SWDF.

\begin{figure}
\centering
\includegraphics[width=4cm]{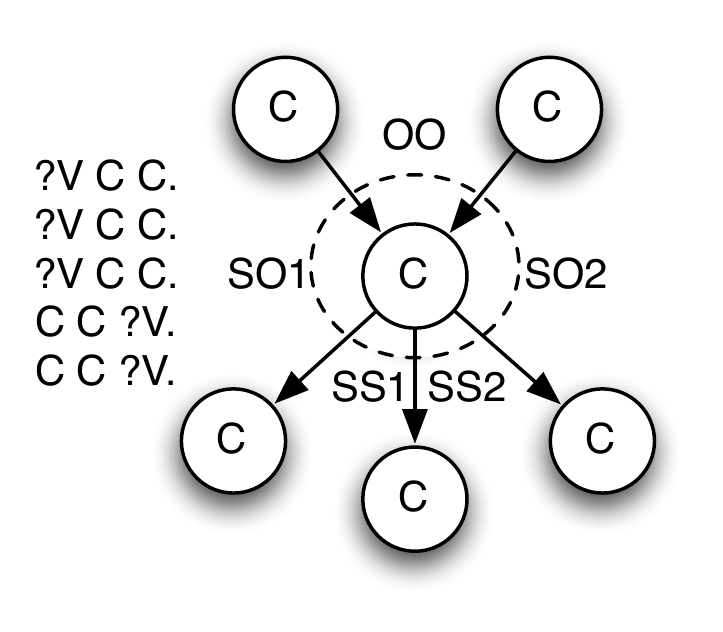}
\caption{Join count example.}
\scriptsize
\label{fig:joinexample}
\end{figure}

The cornerstone of efficient SPARQL evaluation is optimizing the planification of join order execution. Thus, we need to count the number of joins appearing on each query and their types. We can define the join operation as a conjunction of two triple patterns, where both have one variable in common. This leads to six types of joins depending on which position the common variable appears in each pattern: \emph{Subject-Subject}, \emph{Predicate-Predicate}, \emph{Object-Object}, \emph{Subject-Predicate}, \emph{Subject-Object} and \emph{Predicate-Object}\footnote{henceforth referred to using their capital letters}. The SPARQL specification does not determine in which order the joins shall be performed since the result is equivalent due to the commutative property. It is the task of the query evaluation engine to decide the final order of the processing. Given a single query, there is not a unique way of taking the groups, hence, the count of join types varies. In figure~\ref{fig:joinexample} we can see an example of the different join possibilities among 5 graph patterns and one variable ?V, being C non-relevant constants. In this case one of the joins is redundant, and depending on which one we leave out, the join count for each type will differ.

\begin{figure}
\hfill \includegraphics[width=6cm]{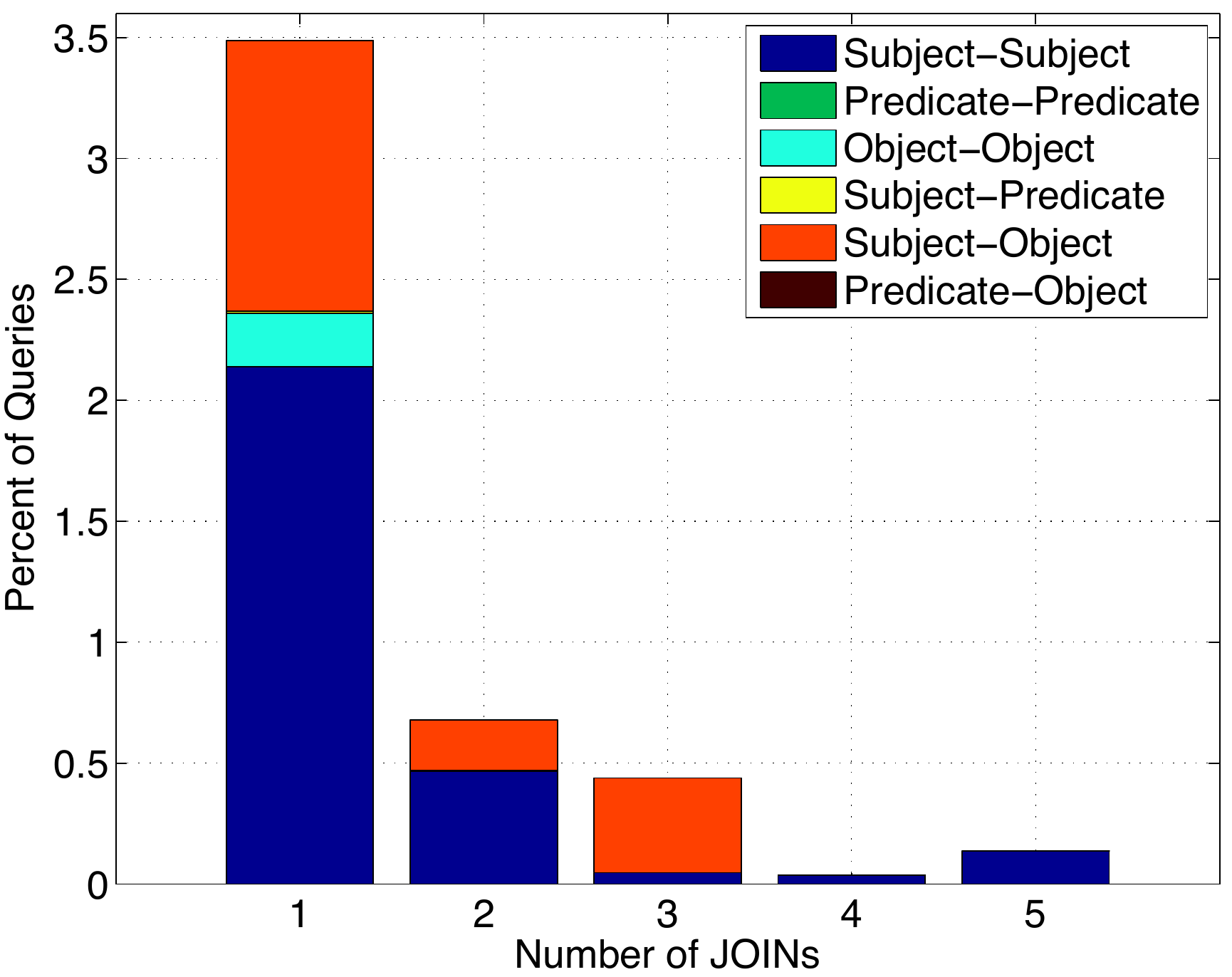}

\hfill \includegraphics[width=7cm]{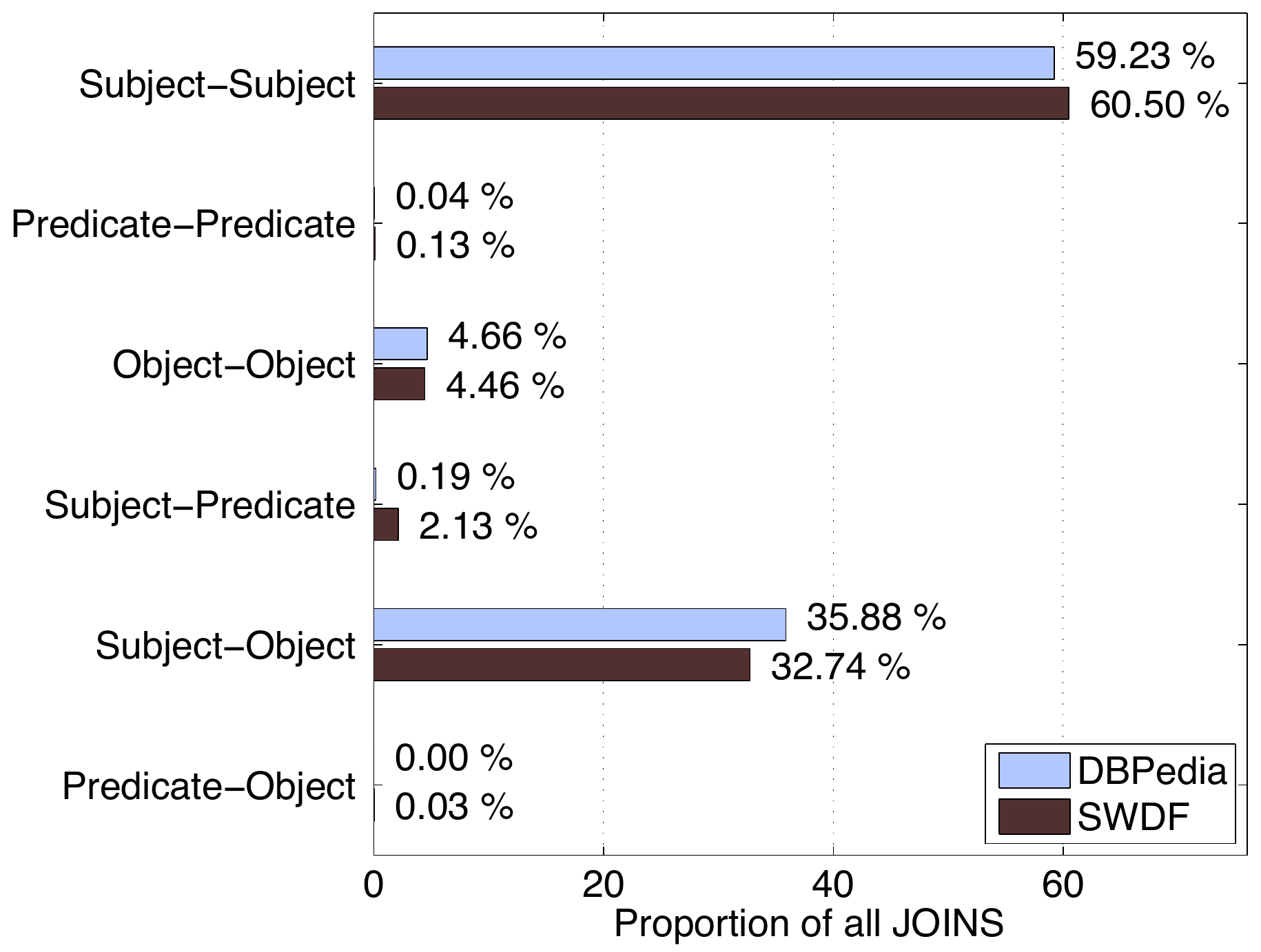}
\caption{Percentage of queries from DBPedia including different join types.}
\scriptsize
\label{fig:joins}
\end{figure}

We propose counting joins by first grouping the same type ones (SS, PP, OO) and then those of different type (SP, SO, PO). This is a simple and consistent method of evaluating join types regardless of the query evaluation engine.
Figure~\ref{fig:joins} shows the results of our join counting method applied to the log data set. We noticed that $2.66\%$ of the total queries in DBPedia have a single join, $0.75\%$ have two joins, and this percentage gradually decreases with a maximum of $10$ joins in a query. In all, $4.25\%$ of the queries have at least one join (see figure~\ref{fig:constructs}). It is also remarkable that the most common type of join is SS, followed by SO and OO.

Another interesting hypothesis posed on previous works (and assumed as true) is that SPARQL graph patterns are typically star-shaped or include long-chains~\cite{Neumann2010}. We propose an analysis to provide empirical evidence to check whether these two statements are valid. To do so we construct a directed graph for each query (similar to the one used in figure~\ref{fig:joinexample}) and calculate its longest path. We discovered that $98\%$ of both DBPedia and SWDF queries had a length of just $1$, $1.8\%$ had $2$, and very few queries had up to $5$ jumps. Thus, there conclude that graph pattern do include chains, but at least in our data sets they are very scarce.

\begin{table}
\centering
\scriptsize
\begin{tabular}{|c|r|r|} \hline
Pattern & DBPedia & SWDF \\\hline
1 0  & 66.512\% & 97.463\%\\\hline
3 0 0 0  & 26.683\% & 0.106\%\\\hline
2 0 0  & 3.773\% & 1.024\%\\\hline
1 1 0  & 1.371\% & 0.482\%\\\hline
5 0 0 0 0 0  & 0.701\% & 0.010\%\\\hline
2 1 0 0  & 0.313\% & 0.432\%\\\hline
3 1 0 0 0  & 0.195\% & 0.040\%\\\hline
4 0 0 0 0  & 0.179\% & 0.020\%\\\hline
6 0 0 0 0 0 0  & 0.107\% & 0.001\%\\\hline
8 0 0 0 0 0 0 0 0  & 0.068\% & 0.000\%\\\hline
6 1 0 0 0 0 0 0  & 0.029\% & 0.001\%\\\hline
<Others> & 0.07\% & 0.420\% \\\hline
\end{tabular}
\caption{Pattern graph out degree serialization.}
\label{tab:outdegree}
\end{table}

In order to characterize whether pattern graphs have star shapes, we propose using a serialization of the out-degree of each node of the graph in decreasing order. Star-shaped graphs will have a central node with a high out-degree, and several leaf nodes with null out-degree (For instance: \texttt{3 0 0 0}). Then we count the frequency of each degree pattern as shown in table~\ref{tab:outdegree}. We see that the most common is the most simple one with one single triple, which might be considered a trivial star and chain. If we keep browsing the rest degree patterns, we observe that there is a big proportion of appearances (more than the $99.93\%$) of almost-star-shaped graph structures between $3$ and $9$ nodes. 

\section{Conclusions and Future Work}
\label{s:cfw}
We noticed that previous analysis of SPARQL queries were not deep enough to take scientifically-supported design decissions when devising RDF stores. Hence, we carried out a study of real-world SPARQL queries in order to understand how real users construct them. We expect our results to be valuable to RDF store designers, specially in the tasks of query evaluation planification and index construction.
We consider our study to be fairly representative, since we have analyzed a large RDF data set log such as DBPedia, and then we have contrasted those results with SWDF. 

We conclude that most queries are simple, \emph{i.e.}, $66.41\%$ of DBPedia queries and $97.25\%$ of SWDF just contain a single triple pattern. However, there are many examples of queries including expensive SPARQL operations, like \texttt{UNION}, \texttt{OPTIONAL} and joins. The percent of queries using join ranges from $2.19\%$ to $4.25\%$, and they are typically of the types SS($\sim60\%$), SO($\sim35\%$) and OO($\sim4.5\%$). We also detected that most queries ($99.97\%$) have a star-shaped graph pattern, and the chains in $98\%$ of the queries have length one, with the longest path having a length of five. 

In future works, we plan to extend our work to other query logs to assess which of our observed behaviours are generalizable, and which are more domain-dependent.
We will also study how the different query clauses affect final solutions, paying special attention to performance. Then, we will be able to provide helpful advices to practitioners on how to leverage our results to improve real-world systems.

\section{Acknowledgments}
Partially funded by the MICINN (TIN2009-14009-C02-02) and the Millennium Institute for Cell Dynamics and Biotechnology (ICDB) (Grant ICM P05-001-F). The second author is granted by a fellowship from Erasmus Mundus, the Regional Government of Castilla y Leon (Spain) and the European Social Fund.

%
%
%
\end{document}